\documentclass[
  10pt,  
  aps,  
  amsmath,
  amssymb,
  twocolumn,
  prb,
  superscriptaddress,
  floatfix
]{revtex4} 

\usepackage{bm}
\usepackage{dsfont}
\usepackage{graphicx}
\usepackage{pifont}
\usepackage{textcomp}
\usepackage{gensymb}
\usepackage{accents}
\usepackage{upgreek}
\usepackage{array}
\usepackage{tabularx}
\usepackage{color}
\usepackage{hhline}
\usepackage{multirow}
\usepackage{mathtools}
\usepackage{natbib}
\usepackage{hyperref}

\newcommand{\be}{\begin{equation}}
\newcommand{\e}{\end{equation}}
\newcommand{\beml}{\begin{subequations}}
\newcommand{\eml}{\end{subequations}}
\newcommand{\beq}{\begin{eqnarray}}
\newcommand{\eq}{\end{eqnarray}}
\newcommand{\ba}{\begin{array}}
\newcommand{\ea}{\end{array}}
\newcommand{\bpm}{\begin{pmatrix}}
\newcommand{\epm}{\end{pmatrix}}
\newcommand{\bc}{\begin{cases}}
\newcommand{\ec}{\end{cases}}

\newcommand{\bb}{\boldsymbol}

\DeclareMathOperator{\sign}{sign}

\begin{document}

\title{Quartic asymmetric exchange for two-dimensional ferromagnets with trigonal prismatic symmetry}

\author{I.\,A.~Ado}
\affiliation{Radboud University, Institute for Molecules and Materials, 6525 AJ Nijmegen, The Netherlands}
\affiliation{Institute for Theoretical Physics, Utrecht University, 3584 CC Utrecht, The Netherlands}

\author{Gulnaz Rakhmanova}
\affiliation{ITMO University, Faculty of Physics, Saint-Petersburg, Russia}

\author{Dmitry A. Zezyulin}
\affiliation{ITMO University, Faculty of Physics, Saint-Petersburg, Russia}

\author{Ivan Iorsh}
\affiliation{ITMO University, Faculty of Physics, Saint-Petersburg, Russia}

\author{M.~Titov}
\affiliation{Radboud University, Institute for Molecules and Materials, 6525 AJ Nijmegen, The Netherlands}

\begin{abstract}
We suggest a possible origin of noncollinear magnetic textures in ferromagnets (FMs) with the $D_{3h}$ point group symmetry. The suggested mechanism is different from the Dzyaloshinskii-Moriya interaction (DMI) and its straightforward generalizations. The considered symmetry class is important because a large fraction of all single-layer intrinsic FMs should belong to it. In particular, so does a monolayer Fe\textsubscript{3}GeTe\textsubscript{2}. At the same time, DMI vanishes identically in materials described by this point group, in the continuous limit. We use symmetry analysis to identify the only possible contribution to the free energy density in two dimensions that is of the fourth order with respect to the local magnetization direction and linear with respect to its spatial derivatives. This contribution predicts long-range conical magnetic spirals with both the average magnetization and the ``average chirality'' dependent on the spiral propagation direction. We relate the predicted spirals to a recent experiment on Fe\textsubscript{3}GeTe\textsubscript{2}. Finally, we demonstrate that, for easy-plane materials, the same mechanism may stabilize bimerons.
\end{abstract}

\maketitle 
\textbf{Introduction.} Isolation of graphene in 2004~\cite{novoselov2004electric,novoselov2005two} attracted remarkable interest to the field of purely two-dimensional (2D) materials, which has been growing to date. It is both fundamental interest and potential applications that drive the research in this field~\cite{Novoselov2005, zhang2005experimental, PhysRevB.93.134407, cortie2020two}. Of particular importance for the applications is the fact that low-dimensional systems can be tuned in a much more effective way than their bulk counterparts~\cite{liang2018inducing, burch2018magnetism}. It is also important that one can combine properties of different 2D materials by stacking them in heterostructures~\cite{geim2013van, novoselov20162d, gibertini2019magnetic}.

Potential applications of heterostructures and monolayers include, among others, spin-based computer logic and ways to store information~\cite{soumyanarayanan2016emergent, gong2019two}. In particular, it is assumed that noncollinear magnetic textures, like skyrmions and miniature domain walls, might become the basis for future memory devices~\cite{parkin2008magnetic, parkin2015memory, everschor2018perspective}. However, for more than a decade, no atomically thin intrinsic magnets have been realized in experiments. This happened only in 2017 when magnetic order was reported in two-dimensional van der Waals materials Cr\textsubscript{2}Ge\textsubscript{2}Te\textsubscript{6}~\cite{gong2017discovery} and CrI\textsubscript{3}~\cite{huang2017layer}. Soon they were accompanied by a metallic itinerant ferromagnet Fe\textsubscript{3}GeTe\textsubscript{2}~\cite{deng2018gate,fei2018two} (FGT).

Recently, spin spirals were reported in thin multilayers of FGT~\cite{meijer2020chiral}, and N\'{e}el-type skyrmions were observed in two different heterostructures based on this material~\cite{wu2020neel, PhysRevB.103.104410}. It is however interesting that noncollinear magnetic order in pure FGT cannot be explained by the Dzyaloshinskii-Moriya interaction~\cite{Dzyaloshinsky1958,Moriya1960} (DMI). The reason for this is the following. Bulk FGT has an inversion symmetry center, and thus smooth noncollinear textures cannot originate in the contributions to the free energy that are associated with DMI. Monolayer FGT, on the other hand, does lack the inversion symmetry. But its point group $D_{3h}$ is still so symmetric that any contribution to the free energy density of the form $n_i\nabla_j n_k$ can affect magnetic order only at the sample boundaries ($\bb n$ here is the unit vector of the local magnetization direction). This fact was coined in Ref.~[\onlinecite{PhysRevB.99.104422}] and repeated in a recent paper~\cite{PhysRevB.102.060402} with an illustrative title \textit{``Elusive Dzyaloshinskii-Moriya interaction in monolayer Fe\textsubscript{3}GeTe\textsubscript{2}''.} Some of us also mentioned this in Ref.~[\onlinecite{PhysRevB.101.161403}].

In addition to monolayer FGT, the group $D_{3h}$ describes many other 2D ferromagnets (FMs). For example, some transition metal dichalcogenides (TMDs), when thinned down to a single layer, are predicted to be intrinsically magnetic~\cite{ataca2012stable}. Typically, 2D TMDs are formed in either 1T or 2H phases~\cite{ataca2012stable,memarzadeh2021role}, and the latter phase is characterized by $D_{3h}$. Another large group of magnetic monolayers, for which the 2H phase (and the $D_{3h}$ symmetry) is often favourable, is transition metal dihalides~\cite{jiang2021recent}. Recently predicted 2D chromium pnictides~\cite{PhysRevB.102.024441} that are half-metallic FMs with very high Curie temperatures are described by the point group $D_{3h}$ as well. Overall, $D_{3h}$, which is the group of symmetries of a triangular prism, is an important group in the field of intrinsic 2D magnetism. In this paper, we introduce a possible origin of smooth noncollinear magnetic textures in materials descibed by this point group.

\textbf{Symmetry analysis.} The \textit{elusive} nature of DMI in such materials is characterized by vanishing antisymmetric contributions $n_i\nabla_j n_k-n_k\nabla_j n_i$ to the free energy density. At the same time, similar symmetric terms are represented by full derivatives $\nabla_j(n_i n_k)$ and therefore can be relevant only close to the edges of the sample. Thus, away from the edges, terms that are quadratic with respect to~$\bb n$ do not contribute to formation of smooth noncollinear magnetic textures at all. As such textures are characterized by small spatial derivatives of magnetization, it is worthwhile to consider contributions to the free energy density that are quartic with respect to $\bb n$ and linear with respect to $\nabla_i n_p$. Namely we would like to study terms of the form $n_i n_j n_k\nabla_l n_p$. We call them the ``quartic asymmetric exchange'' terms by analogy with DMI. Physically, they can correspond, for example, to interactions between four spins~\cite{rybakov2020antichiral, ado2020non}.

Let us use the standard symmetry analysis~\cite{Authier2003book,PhysRevLett.119.127203} to identify all quartic asymmetric exchange terms allowed in~$D_{3h}$. This group contains a 3-fold rotation around the $z$ axis, a mirror symmetry with respect to the $xy$ plane, and three 2-fold rotations around the axes at the angles $0$, $\pm2\pi/3$ in the $xy$ plane. Quartic contributions
\be
\label{quartic_gen}
I_{\mathcal D}=\sum_{ijklp}{\mathcal D_{ijklp}\cdot n_i n_j n_k\nabla_l n_p}
\e
to the free energy density should remain invariant under the transformation
\be
\label{symmetry_analysis}
\mathcal D_{i'j'k'l'p'}=\sum_{ijklp}{\mathcal D_{ijklp}\cdot g_{ii'}g_{jj'}g_{kk'}g_{ll'}g_{pp'}}
\e
for every group element $g$. By applying all generators of $D_{3h}$ to Eq.~(\ref{symmetry_analysis}), we find that in this group there are precisely seven distinct invariant quartic contributions of the form of Eq.~(\ref{quartic_gen}). We collect them in Table~\ref{table}.

Surprisingly, these invariants can be combined to form five different expressions represented by full derivatives (see also the caption of Table~\ref{table}). Therefore, up to boundary terms, five invariants out of seven are not independent. We can choose, e.\,g.,
\begin{subequations}
\begin{gather}
\label{w_D}
w_\parallel=n_x\left(n_x^2-3n_y^2\right)\left(\nabla_x n_x+\nabla_y n_y\right), \\
w_\perp=n_x\left(n_x^2-3n_y^2\right)\nabla_z n_z
\end{gather}
\end{subequations}
as the only independent ones. Moreover, in a 2D system $\nabla_z\equiv 0$, and the second term, $w_\perp$, can be also disregarded. Hence, if the effects of boundaries are negligible, any 2D FM with the $D_{3h}$ point group symmetry allows only a single quartic asymmetric exchange term: $w_\parallel$. This is in stark contrast to the situation with the quadratic contributions to the free energy density that are linear with respect to $\nabla_i n_p$. Roughly speaking, half of all such terms allowed by symmetry are independent of the others.

\begin{figure}[t!]
\includegraphics[width=0.9\columnwidth]{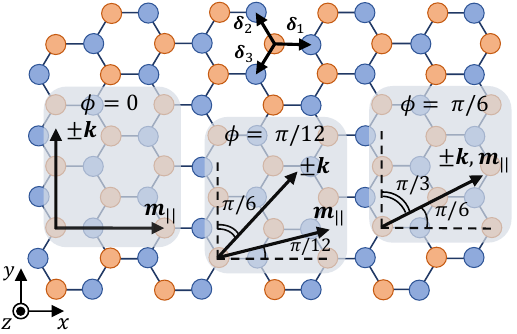}
\caption{Hexagonal lattice typical for many materials with the trigonal prismatic symmetry. Panels show three particular directions of the spiral wave vector $\bb k$ (up to a sign) and the in-plane component $\bb m_\parallel$ of the vector $\bb m$. The latter corresponds to the average magnetization direction and can be controlled by an external magnetic field. Note that directions of both $\bb k$ and $\bb m_\parallel$ are set by the angle $\phi$ (see Eqs.~(\ref{ansatz_m}),~(\ref{k_min})).}
\label{fig::spirals}
\end{figure}

It is useful to relate the structure of $w_\parallel$ to the lattice geometry of a typical 2D crystal described by the point group $D_{3h}$. One can notice that
\be
\label{w_D_symm}
w_\parallel\propto(\bb n\cdot\bb\delta_1)(\bb n\cdot\bb\delta_2)(\bb n\cdot\bb\delta_3)\left(\nabla_x n_x+\nabla_y n_y\right),
\e
where $\bb\delta_i$ represent the nearest neighbour vectors. These vectors or their opposite make the angles $0$, $\pm2\pi/3$ with the positive $x$ axis (see the top part of Fig.~\ref{fig::spirals}). They also correspond to the three armchair directions of a typical hexagonal lattice generated by $D_{3h}$. Using Eq.~(\ref{w_D_symm}), one can obtain a classical Heisenberg model for $w_\parallel$. For a site with the spin $\bb S$, we have
\be
\label{Heisenberg}
w_{\parallel,H}\propto(\bb{S}\cdot\bb{\delta}_{1})(\bb{S}\cdot\bb{\delta}_{2}) (\bb{S}\cdot\bb{\delta}_{3}) \sum_{i} (\bb{S}_i\cdot\bb{\delta}_{i}),
\e
where the spins $\bb{S}_i$ are the nearest neighbours of $\bb{S}$.

We note that the effective interaction of Eq.~(\ref{Heisenberg}) is fundamentally different from the recently proposed~\cite{PhysRevB.99.184430, brinker2019chiral, PhysRevB.101.024418} interactions of the form $(\bb{S}_i\times \bb{S}_j) (\bb{S}_k\cdot \bb{S}_l)$. In the continuous limit, the latter are represented by higher-order terms with respect to the gradients of magnetization direction. Thus, for smooth noncollinear textures, the interaction of Eq.~(\ref{Heisenberg}) is potentially the leading one. At the same time, for textures varying on the scale of the lattice spacing, this is not necessarily the case.

\textbf{Spin spirals.} Now let us find out whether the quartic term $w_\parallel$ can stabilize spin spirals observed in FGT. In order to do this, we consider a conical ansatz
\be
\label{ansatz_gen}
\bb n(\bb r)=\bb m\cos{\alpha}+\left[\bb m_\theta\cos{(\bb k\bb r)}+\bb m_\phi\sin{(\bb k\bb r)}\right]\sin\alpha
\e
which parameterizes the transition from a collinear state, $\sin\alpha=0$, to a helix, $\cos\alpha=0$ (if $\bb k\neq 0$). Here
\begin{subequations}
\label{ansatz}
\begin{gather}
\label{ansatz_m}
\bb m=(\cos{\phi}\sin{\theta},\sin{\phi}\sin{\theta},\cos{\theta}),\\
\bb m_\theta=(\cos{\phi}\cos{\theta},\sin{\phi}\cos{\theta},-\sin{\theta}),\\
\bb m_\phi=(-\sin{\phi},\cos{\phi},0)
\end{gather}
\end{subequations}
is a standard basis in spherical coordinates. Vectors $\bb m_\theta$ and $\bb m_\phi$ correspond to oscillations, while $\bb m$ points in the direction of the average magnetization.

\begingroup
\begin{table}
\begin{center}
{\renewcommand{\arraystretch}{1.35}
\begin{tabular}{c}
\hline\hline
$I_1=n_x\left(n_x^2-3n_y^2\right)\left(\nabla_x n_x+\nabla_y n_y\right)\equiv w_\parallel$ \\
$I_2=n_x\left(n_x^2-3n_y^2\right)\nabla_z n_z\equiv w_\perp$ \\
\hline
$I_3=n_y\left(n_y^2-3n_x^2\right)\left(\nabla_x n_y-\nabla_y n_x\right)$ \\
$I_4=\left(n_x^2+n_y^2\right)\left[\nabla_x\left(n_y^2-n_x^2\right)+2\nabla_y\left(n_x n_y\right)\right]$ \\
$I_5=n_z^2\left[\nabla_x\left(n_y^2-n_x^2\right)+2\nabla_y\left(n_x n_y\right)\right]$ \\
$I_6=n_z\left(n_y^2-n_x^2\right)\nabla_x n_z+2n_x n_y n_z\nabla_y n_z$ \\
$I_7=n_z\left(n_y^2-n_x^2\right)\nabla_z n_x+2n_x n_y n_z\nabla_z n_y$ \\
\hline\hline
\end{tabular}
}
\end{center}
\caption{\label{table}All fourth-order contributions to the free energy density that are linear with respect to spatial derivatives of~$\bb n$. Using the constraint $\bb n^2=1$, one can show that the following five combinations correspond only to boundary terms: \mbox{$I_1+I_3$}, $I_1+5I_3/3+I_4/2$, $I_2-3I_7$, $I_4+I_5$, $I_5+2I_6$ (i.\,e., they can be expressed by full derivatives). Therefore, away from the edges, only two invariants out of seven can be chosen as independent. We choose the pair $I_1$, $I_2$ as such. Note that, in sufficiently small samples, boundary effects cannot be ignored, and all seven invariants should be taken into account.}
\end{table}
\endgroup

We substitute this ansatz into a model that accounts for the symmetric exchange ($A$), magnetic anisotropy ($K$), and the quartic asymmetric exchange ($\mathcal D$),
\be
\label{free_energy_def}
f=A\left[(\nabla_x\bb n)^2+(\nabla_y\bb n)^2\right]+ Kn_z^2+8\mathcal Dw_\parallel,
\e
and average the total free energy $F=\int{d^2r f}$ over a large volume. Oscillating terms with $\bb k\bb r$ then vanish, and the averaged density $\langle f \rangle$ becomes a quadratic function of the wave vector $\bb k$. Therefore, we straightforwardly minimize it with respect to $\bb k$ and find:
\begin{multline}
\label{free_energy_k_min}
\frac{A}{\mathcal D^2}\langle f \rangle=-\frac{9}{64}\left(\sin{\alpha}+5\sin{3\alpha}\right)^2\sin^4{\theta}\cos^2{\theta}\\+
\frac{AK}{2\mathcal D^2}\left(\sin^2{\theta}\sin^2{\alpha}+2\cos^2{\theta}\cos^2{\alpha}\right),
\end{multline}
where the combination $AK/\mathcal D^2$ is dimensionless, and further minimization with respect to $\theta$ and $\alpha$ is required. The wave vector that corresponds to the minimum is expressed as 
\be
\label{k_min}
\begin{pmatrix}
k_x\\
k_y
\end{pmatrix}
=-\frac{3\mathcal D}{2A}\left(5\cos^2{\alpha}-1\right)\sin^2{\theta}\cos{\theta}
\begin{pmatrix}
\sin{2\phi}\\
\cos{2\phi}
\end{pmatrix}.
\e

Before we proceed with the minimization, it is interesting to note that states described by Eq.~(\ref{free_energy_k_min}) are degenerate with respect to $\phi$. In other words, their free energy does not depend on the azimuthal angle of ``the average magnetization vector'' $\bb m$. At the same time, for spiral-like textures with a finite wave vector, the angles between $\bb k$ and $\bb m$ are different for different values of $\phi$:
\be
\label{m_k_angle}
\bb k\cdot\bb m\propto\sin{3\phi},
\e
as it follows from Eqs.~(\ref{ansatz_m},\ref{k_min}). Thus, by controlling the direction of the average magnetization (vector $\bb m$), one should also be able to control the propagation direction of the spiral. Such control can be achieved by an application of a small external magnetic field. The latter couples only to $\bb m$, therefore it can be used to set the desired value of $\phi$. As we see from Eq.~(\ref{m_k_angle}), when the in-plane component $\bb m_\parallel$ of the vector $\bb m$ lies along one of the armchair directions of the lattice ($\phi=0,\,2\pi/3,\,4\pi/3$), then $\bb k$  is orthogonal to it. When $\bb m_\parallel$ is along a zigzag direction ($\phi=\pi/6,\,5\pi/6,\,3\pi/2$), then $\bb k$ (or $-\bb k$) points in the same direction (see the bottom part of Fig.~\ref{fig::spirals}). We wonder whether such control of the wave vector direction can be realized experimentally.

\begin{figure}[b!]
\includegraphics[width=0.99\columnwidth]{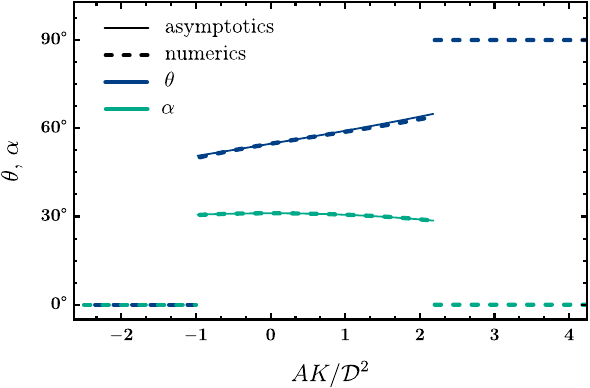}
\caption{Acute angles that correspond to the global minimum of the functional given by Eq.~(\ref{free_energy_k_min}). In fact, for $AK/\mathcal D^2\gtrsim 2.18$, the configuration with $\sin^2{\alpha}=1$, $\sin^2{\theta}=0$ has the same energy as the one with $\sin^2{\alpha}=0$, $\sin^2{\theta}=1$ shown here. However, both configurations describe the same collinear state with $\bb n$ lying in the $xy$ plane (i.\,e., with $\pm\bb n=\bb m=\bb m_\parallel$). Solid curves correspond to the expressions of Eqs.~(\ref{spiral_angle}). Blue (green) colour represent $\theta$ ($\alpha$).}
\label{fig::theta_alpha}
\end{figure}

Let us proceed with the consideration of the functional defined by Eq.~(\ref{free_energy_k_min}). Perturbative analysis with respect to small $AK/\mathcal D^2$ provides almost a perfect fit for its minimum in the entire range of parameters. States with $\bb k\neq 0$ can exist when $-0.98\lesssim AK/\mathcal D^2\lesssim 2.18$, and for such states we find
\begin{subequations}
\label{spiral_angle}
\begin{gather}
\label{spiral_angle_theta}
\sin^2{\theta}=\frac{2}{3}+\frac{9}{128}\frac{AK}{\mathcal D^2}+\dots,\\
\label{spiral_angle_alpha}
\sin^2{\alpha}=\frac{4}{15}-\left(\frac{9}{32\sqrt{10}}\frac{AK}{\mathcal D^2}\right)^2+\dots,
\end{gather}
\end{subequations}
where only the leading and the subleading order terms are shown. For other values of the parameter $AK/\mathcal D^2$, the state is collinear: out-of-plane for $K<0$ and in-plane for $K>0$ (see Fig.~\ref{fig::theta_alpha}). This resembles a typical situation with magnetic textures determined by DMI: when the absolute value of the DMI strength $D$ exceeds some critical value $D_{c}\propto\sqrt{A|K|}$, the system is found in a helical ground state, while for $|D|<D_{c}$ the uniform magnetization is favoured~\cite{PhysRevLett.105.157201, PhysRevB.97.064403}.

Helical states described by Eqs.~(\ref{ansatz_gen}, \ref{ansatz}, \ref{k_min}, \ref{spiral_angle}) are in a reasonable agreement with the spirals found in FGT. For $\phi=0$ and large enough $|\mathcal D|$, we have a helical texture with finite $\bb k$ pointing along the $y$ direction. The components $n_y$ and $n_z$ of this texture oscillate with a phase difference of $\pi/2$ (see Eqs.~(\ref{ansatz_gen}),~(\ref{ansatz})). This is precisely what has been reported in Ref.~[\onlinecite{meijer2020chiral}]. In addition, however, we have an oscillating $n_x$ component, with a slightly smaller amplitude that is, basically, equal to $\cos{\theta}$. It is not clear whether this is a crucial disagreement with the experiment or something that was not seen in it. We should also note that $x$ and $y$ axes in Ref.~[\onlinecite{meijer2020chiral}] describe the coordinates of detectors, not the  crystal axes. Hence the~$x$ component of magnetization in that paper can indeed correspond to our~$n_y$.

\textbf{Chirality.} One can distinguish between magnetic textures with different chiralities (handednesses) by computing the quantity~\cite{rybakov2020antichiral}
\be
\label{chirality_def}
\rho=\bb n\cdot[\bb\nabla\times\bb n].
\e
Texture that can be superposed on its mirror image is called achiral, and in this case, $\rho=0$. If, on the other hand, $\rho=const$, then the texture is called chiral, and it cannot be superposed on its reflection. We also say that the sign of $\rho$ determines one of the two possible handednesses of a texture.

It is interesting that the quartic asymmetric exchange does not come with a preferred chirality. Moreover, the ``average chirality'' of the conical spirals under consideration can be externally controlled. To see this, we employ Eqs.~(\ref{ansatz_gen}, \ref{ansatz}, \ref{free_energy_k_min}, \ref{k_min}, \ref{chirality_def}) and observe that for our spirals
\begin{equation*}
	\rho=\frac{\mathcal D}{A}\left[\Upsilon_0\sin{3\phi}+\Upsilon_1\cos{(\bb k\bb r \hspace{-0.1ex} + \hspace{-0.1ex} 3\phi)}+\Upsilon_2\cos{(\bb k\bb r \hspace{-0.1ex} - \hspace{-0.1ex} 3\phi)}\right],
\end{equation*}
where $\Upsilon_i$ are dimensionless functions of the parameter $AK/\mathcal D^2$. Full expressions for $\Upsilon_1$ and $\Upsilon_2$ are not important for us here, while for $\Upsilon_0$ we can write
\be
\label{Upsilon_0}
\Upsilon_0=\frac{3}{2}\left(5\cos^2{\alpha}-1\right)\sin^2{\alpha}\sin^3{\theta}\cos{\theta},
\e
where $\alpha$ and $\theta$ correspond to Eqs.~(\ref{spiral_angle}). After averaging over the period or over a large volume, we find
\begin{equation}
	\langle\rho\rangle=\frac{\mathcal D}{A}\Upsilon_0\sin{3\phi}.
\end{equation}

It can be seen from Fig.~\ref{fig::theta_alpha} and Eq.~(\ref{Upsilon_0}) that $\Upsilon_0$ does not change its sign (and approximately equals 0.3) when \mbox{$-0.98\lesssim AK/\mathcal D^2\lesssim 2.18$}. Therefore, for a fixed $\mathcal D\neq 0$, the average chirality $\langle\rho\rangle$ of the spiral is generally finite and can be fully controlled, e.\,g. by a small external magnetic field. An interesting discussion of magnetic textures with $\rho\neq 0$, $\langle\rho\rangle=0$ can be found in Ref.~[\onlinecite{rybakov2020antichiral}].

\textbf{Skyrmions and bimerons.} So far we have analyzed the ``spiral region'' of the model's phase portrait, \mbox{$-0.98\lesssim AK/\mathcal D^2\lesssim 2.18$}. Let us now turn to the opposite case, when collinear states are preferred over spin spirals. In such a case, skyrmions and bimerons can in principle exist as metastable transitions from one collinear state to another. Skyrmions may potentially be realized in easy-axis magnets, $K<0$ in Eq.~(\ref{free_energy_def}), while materials with an easy-plane anisotropy, $K>0$, might host bimerons (which are the in-plane skyrmions~\cite{zhang2015magnetic, PhysRevLett.119.207201,PhysRevB.99.060407}). Note that FGT is an easy-axis magnet~\cite{PhysRevB.93.134407}.

First, we observe that, in fact, circular skyrmions cannot be stabilized by the quartic exchange $w_\parallel$. For a standard ansatz $\bb n(\bb r)=\left(\cos{\Phi}\sin{\Theta},\sin{\Phi}\sin{\Theta},\cos{\Theta}\right)$, with $\Phi=Q \phi+\delta$ and $\Theta=\Theta(r)$, integration over $\phi$ nullifies $w_\parallel$ if $Q$ is an integer. At the same time, there exist predictions of skyrmions with more complex axial symmetry~\cite{mcgrouther2016internal}, including trigonal~\cite{pepper2018skyrmion, behera2018size}. Such symmetry is natural for $D_{3h}$, and we have checked that indeed, for ``trigonal skyrmions'', $w_\parallel$ is generally finite after the angle integration. Nevertheless, we do not think that this can explain skyrmions observed in the experiments of Refs.~[\onlinecite{wu2020neel}],~[\onlinecite{PhysRevB.103.104410}]. This is however not unexpected because the spatial symmetries of heterostructures studied in these works are anyway not described by $D_{3h}$.

\begin{figure}[t!]
\includegraphics[width=0.95\columnwidth]{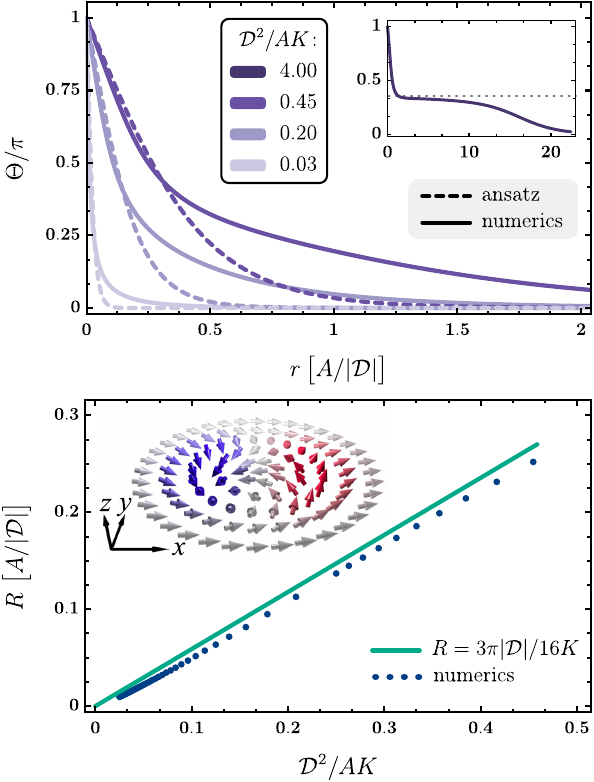}
\caption{Top panel: numerical solutions of Eq.~(\ref{Lagrange}) compared with the ansatz of Eq.~(\ref{ansatz_dw}) for different values of the parameter $\mathcal D^2/AK$. It is assumed that $\sin{(3\phi_0+\delta)}=-\sign{\mathcal D}$. The inset illustrates how two domain walls match when $\cos{\Theta(r)}\approx1/\sqrt{5}$. Bottom panel: bimeron radius $R$ calculated numerically and estimated analytically. The inset shows a bimeron with $\delta=\pi/2$, $\phi_0=0$, $Q=1$ (see also Eqs.~(\ref{bimeron_def})). We measure $r$ and $R$ in units of  $A/|\mathcal D|$. We note that the numerical curves shown in the top panel are strongly nonlinear around $r=0$. Therefore, one cannot replace the ansatz of Eq.~(\ref{ansatz_dw}) with a linear function of $r$.}
\label{fig::profiles_R}
\end{figure}

The situation with bimerons is different. For $K>0$, a~collinear state is more energetically favorable than a spin spiral when \mbox{$\mathcal D^2\lesssim AK/2.18\approx 0.46 AK$}. It turns out that, in this parameter region, a bimeron can indeed exist as a transition between two collinear in-plane states. To demonstrate this, let us consider a parametrization
\begin{subequations}
\label{bimeron_def}
\begin{gather}
\label{bimeron_def1}
\bb n(\bb r)=\hat{\mathcal R}_z[\phi_0]\left(\cos{\Theta},\cos{\Phi}\sin{\Theta},\sin{\Phi}\sin{\Theta}\right),\\
\Phi=Q(\phi+\phi_0)+\delta,\qquad\Theta=\Theta(r),
\end{gather}
\end{subequations}
with the boundary conditions $\Theta(0)=\pi$, $\Theta(\infty)=0$. Here $Q$ is the bimeron's topological charge, and $\hat{\mathcal R}_z[\phi_0]$ denotes the matrix of rotation by an arbitrary angle~$\phi_0$ with respect to $z$. Eqs.~(\ref{bimeron_def}) desribe a bimeron magnetized at $r=\infty$ in the direction set by the polar angle $\phi_0$. The inset of the bottom panel of Fig.~\ref{fig::profiles_R} provides an illustration of such a bimeron with $\delta=\pi/2$, $\phi_0=0$, $Q=1$.

We substitute Eqs.~(\ref{bimeron_def}) with $Q=1$ into Eq.~(\ref{free_energy_def}), integrate over $\phi$, and compute a functional derivative of the result. This brings us to the Euler-Lagrange equation
\begin{align}
\Theta''(r)+\frac{\Theta'(r)}{r}-\frac{\sin{2\Theta(r)}}{2r^2}&-\frac{K\sin{2\Theta(r)}}{4A}\nonumber \\
\label{Lagrange}
-\frac{3\mathcal D\sin{(3\phi_0+\delta)}}{2Ar}\sin^2{\Theta(r)}&\left[5\cos^2{\Theta(r)}-1\right]=0.
\end{align}
It is very similar to the equation that describes skyrmions in the presence of the DMI term $n_z(\bb\nabla\cdot\bb n)-(\bb n\cdot\bb\nabla) n_z$. From Refs.~[\onlinecite{wang2018theory}],~[\onlinecite{PhysRevB.97.064403}], we know that the radial profile of such skyrmions can be very well approximated by a domain wall with two parameters. We can use the approach of these papers to analyze solutions of Eq.~(\ref{Lagrange}) as well.

The only difference between the Euler-Lagrange equation for skyrmions and Eq.~(\ref{Lagrange}) is the presence of the term $5\cos^2{\Theta(r)}-1$ in the latter. At small values of $r$, its effect on the solutions is minimal. But for larger $r$, when $\cos{\Theta(r)}\approx\pm1/\sqrt{5}$, this term becomes important. Therefore, we can expect that the bimeron profile $\Theta(r)$ is a superposition of a domain wall and some additional structure that is relevant at large $r$. Being optimistic, one can hope that at least some properties of $\Theta(r)$ can be captured from the analysis of its domain wall ``component'' alone. It turns out that this is indeed the case.

We employ the ansatz of Ref.~[\onlinecite{wang2018theory}]:
\be
\label{ansatz_dw}
\Theta_{\text{dw}}(r)=2\arctan{\left[\frac{\sinh{\left(R/\Delta\right)}}{\sinh{\left(r/\Delta\right)}}\right]},
\e
where $\Delta$ is the width of the domain wall, and $R$ is the profile radius: $\Theta_{\text{dw}}(R)=\pi/2$. Assuming $R\gg\Delta$ and repeating the considerations of Ref.~[\onlinecite{wang2018theory}], we can estimate the free energy $F$ of this ansatz as
\begin{equation*}
F\approx4\pi\left[A\left(\frac{R}{\Delta}+\frac{\Delta}{R}\right)+\frac{KR\Delta}{2}+\frac{3\pi\mathcal D R}{16}\sin{(3\phi_0+\delta)}\right].
\end{equation*}
Based on this result, we can argue that the minimal energy corresponds to $\sin{(3\phi_0+\delta)}=-\sign{\mathcal D}$. Alternatively, one can see this from the direct minimization of the above expression (using the fact the $R$ should be positive). Either way, we minimize our expression for $F$ with respect to both $R$ and $\Delta$ to obtain
\be
\Delta=\frac{3\pi |\mathcal D|}{16K},\qquad R=\frac{\Delta}{\sqrt{1-K\Delta^2/2A}}.
\e
For $\mathcal D^2/AK\lesssim 0.46$, the square root can be safely ignored, and we are left with $R=\Delta=3\pi |\mathcal D|/16K$.

This result obviously contradicts the initial assumption $R\gg\Delta$. Nevertheless, it works astonishingly well when $r\lesssim R$, as can be seen from Fig.~\ref{fig::profiles_R}. There, we plot numerical solutions of Eq.~(\ref{Lagrange}) which was supplemented with the condition $\sin{(3\phi_0+\delta)}=-\sign{\mathcal D}$. For $\Theta(r)\gtrsim\pi/2$, the ansatz $\Theta_{\text{dw}}(r)$ correctly reproduces the shape of the bimeron (top panel) and allows us to get a good estimate of its radius (bottom panel). Out of curiosity, we also solved Eq.~(\ref{Lagrange}) for $\mathcal D^2/AK=4$ (inset of the top panel). The result looks like a superposition of two domain walls that match at $\cos{\Theta(r)}\approx1/\sqrt{5}$. This is of course by far the ``spiral region'' of our model.

We note that the relation between the bimeron direction $\phi_0$ and its phase~$\delta$ can be more complex than $\sin{(3\phi_0+\delta)}=-\sign{\mathcal D}$. One should perform a thorough numerical analysis to establish such a relation, for every particular combination of $\mathcal D^2/AK$ and $\sign{\mathcal D}$. Once this is done, it is also required to investigate whether the obtained bimeron solution is stable. We checked that, at least for $\mathcal D\sin{(3\phi_0+\delta)}<0$, solutions of Eq.~(\ref{Lagrange}) are strong minimums of the free energy (see Appendix~\ref{appendix1} for a more detailed discussion). In particular this is true for $\sin{(3\phi_0+\delta)}=-\sign{\mathcal D}$. Thus we argue that $w_\parallel$ can indeed stabilize a bimeron (regardless of the assumptions we used to estimate its radius). 

\textbf{Conclusions.} We used symmetry analysis to obtain all contributions to the free energy density of the form $n_i n_j n_k\nabla_l n_p$ that are allowed by the point group $D_{3h}$. There are exactly seven such contributions. Only two of them can be chosen as independent if boundary terms are ignored, and only one of these two does not vanish in a 2D system. We demonstrated that the quartic term $w_\parallel=n_x\left(n_x^2-3n_y^2\right)\left(\nabla_x n_x+\nabla_y n_y\right)$ looks compatible with the spin spirals observed in a recent experiment on FGT. It does not stabilize circular skyrmions, but for FMs with an easy plane, it can stabilize bimerons. We estimated the radius and energy of such bimerons analytically and calculated their profiles numerically by solving the Euler-Lagrange equation. We argue that the quartic asymmetric exchange term $w_\parallel$ introduced in this paper should be considered a potential source of noncollinear magnetic textures in many 2D intrinsic ferromagnets. To investigate the role of this term numerically, one may use the effective Heisenberg model that we derived. We also note that there exist many mechanisms that can produce noncollinear magnetic textures. Here we propose a natural replacement of DMI for systems where the latter is forbidden.

\begin{acknowledgments}
We are grateful to Marcos H. D. Guimar\~{a}es for answering numerous questions about the experiment of Ref.~[\onlinecite{meijer2020chiral}] and to Alexander Rudenko for insights regarding 2D FMs. This research was supported by the JTC-FLAGERA Project GRANSPORT.
\end{acknowledgments}


\bibliographystyle{apsrev4-1}
\bibliography{Bib}

\appendix\onecolumngrid

\section{Dimensionless equations and stability}
\label{appendix1}
We introduce $d=\mathcal D\sin{(3\phi_0+\delta)}$ and $\rho=r|d|/A$. This allows us to rewrite the Euler-Lagrange Eq.~(\ref{Lagrange}) in the dimensionless form
\be
    \label{Lagrange_dims}
    \Theta''(\rho)+\frac{\Theta'(\rho)}{\rho}-\frac{\sin{2\Theta(r)}}{2\rho^2}-\frac{A K}{d^2}\frac{\sin{2\Theta(\rho)}}{4}+\frac{3\sigma}{2\rho}\sin^2{\Theta(\rho)}\left[5\cos^2{\Theta(\rho)}-1\right]=0,
\e
where $\sigma=-\sign{\left[\mathcal D\sin{(3\phi_0+\delta)}\right]}$ and it is assumed that $d\neq 0$. If $d=0$, then $\Theta(r)$ in Eq.~(\ref{Lagrange}) describes a skyrmion profile in the absence of DMI. Such a skyrmion has a zero radius~\cite{wang2018theory}. Sufficient conditions for an extremum are considered with the help of the Jacobi accessory equation~\cite{Gelfand}. It reads
\begin{equation}
    \label{Jacobi_dims}
    h''(\rho)+\frac{h'(\rho)}{\rho}-\left\{\left(\frac{2}{\rho}+\rho\frac{A K}{d^2}\right)\cos{2\Theta(\rho)}+3\sigma[\sin{2\Theta(\rho)}-(5/2)\sin{4\Theta(\rho)}]\right\}\frac{h(\rho)}{2\rho}=0,
\end{equation}
where $\Theta(\rho)$ is a solution of Eq.~(\ref{Lagrange_dims}).

In Eqs.~(\ref{Lagrange_dims}) and (\ref{Jacobi_dims}), $\sigma=1$ or $\sigma=-1$. It is not known a priori which one of these two possibilities is realized. In the main text of this paper we have minimized the bimeron free energy for the domain wall ansatz. According to this analysis, the following condition should hold:
\be
\label{phi_delta}
\sin{(3\phi_0+\delta)}=-\sign{\mathcal D}.
\e
It corresponds to $d=-|\mathcal D|$ and $\sigma=1$. We use the latter fact as a motivation to first solve Eqs.~(\ref{Lagrange_dims}) and (\ref{Jacobi_dims}) for this value of~$\sigma$. The solutions are obtained numerically for several values of the parameter $d^2/AK$. The Jacobi accessory equation is solved with the initial conditions $h(0)=0$, $h'(0)=1$. It turns out that none of the solutions $h(\rho)$ have zeroes different from the one at the origin. Moreover, the second derivative of the free energy with respect to $\Theta'$ is equal to the combination $4\pi A\rho$ that is nonnegative. Hence, the sufficient conditions for a strong minimum are satisfied~\cite{Gelfand}. Therefore, we argue that, for $\mathcal D\sin{(3\phi_0+\delta)}<0$, the ansatz of Eqs.~(\ref{bimeron_def}) describes a strong minimum of the free energy, i.\,e. the bimeron is stabilized by the quartic asymmetric exchange term $w_\parallel$.

\begin{figure}[h!]
\includegraphics[width=0.6\columnwidth]{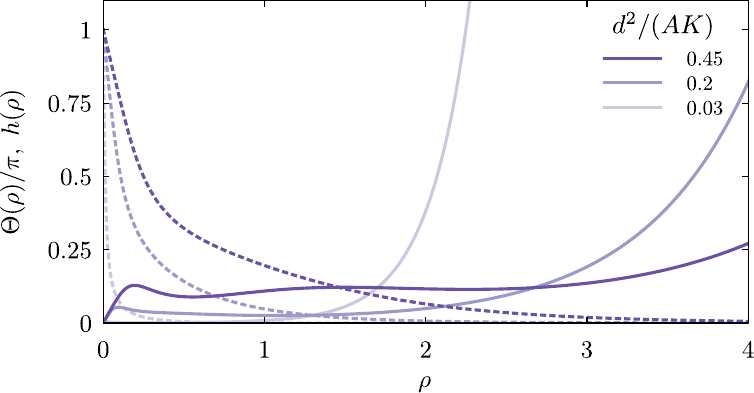}
\caption{Solutions $\Theta(\rho)$ of Eq.~(\ref{Lagrange_dims}) (dashed lines) and the corresponding solutions $h(\rho)$ of Eq.~(\ref{Jacobi_dims}) (solid lines), for $\sigma=1$ and three different values of the parameter $d^2/AK$.}
\label{fig::jacobi}
\end{figure}

In the opposite case, $\sigma=-1$, we are unable to find any solutions of Eq.~(\ref{Lagrange_dims}), using the shooting method. We anticipate that in this case it does not have solutions at all (for the boundary conditions $\Theta(0)=\pi$, $\Theta(\infty)=0$). In other words, if $\mathcal D\sin{(3\phi_0+\delta)}>0$, the bimeron is expected to be unstable. This is in line with the fact that, for the domain wall ansatz, the minimum of the free energy is reached only when $\sigma=1$ (see Eq.~(\ref{phi_delta})).

\section{Relation between $\phi_0$ and $\delta$}
\label{appendix2}
Considerations of the previous section suggest that the condition $\mathcal D\sin{(3\phi_0+\delta)}<0$ ensures the bimeron stability. However, the concrete relation between $\phi_0$ and $\delta$ that corresponds to a global minimum should be obtained for all values of $AK/\mathcal D^2$ and $\sign{\mathcal D}$ by minimizing the bimeron free energy with respect to $\phi_0$ and $\delta$. The result can be more complex than that of Eq.~(\ref{phi_delta}). Nevertheless, it will also be given by a periodic function of $\phi_0$ with a period of $2\pi/3$.

\end{document}